\documentclass[twocolumn,aps,prb,preprintnumbers,amsmath,amssymb]{revtex4} 
\usepackage{graphicx}  
\usepackage{dcolumn}   
\usepackage{bm}        
\usepackage{amssymb}   
\usepackage[dvipsnames,usenames]{color}
\usepackage{color}
\usepackage{comment}
\usepackage{gensymb}
\usepackage{physics}

\begin{document}
\newcommand{\etal}{{\sl et al.}}
\newcommand{\ie}{{\sl i.e.}}
\newcommand{\sto}{SrTiO$_3$}
\newcommand{\lao}{LaAlO$_3$}
\newcommand{\lno}{LaNiO$_3$}
\newcommand{\nith}{Ni$^{3+}$}
\newcommand{\otw}{O$^{2-}$}
\newcommand{\alo}{Al$_2$O$_3$}
\newcommand{\aalo}{$\alpha$-Al$_2$O$_3$}
\newcommand{\xto}{$X_2$O$_3$}
\newcommand{\eg}{$e_{g}$}
\newcommand{\tg}{$t_{2g}$}
\newcommand{\dzt}{$d_{z^2}$}
\newcommand{\dxtyt}{$d_{x^2-y^2}$}
\newcommand{\dxy}{$d_{xy}$}
\newcommand{\dxz}{$d_{xz}$}
\newcommand{\dyz}{$d_{yz}$}
\newcommand{\egp}{$e_{g}'$}
\newcommand{\ag}{$a_{1g}$}
\newcommand{\mub}{$\mu_{\rm B}$}
\newcommand{\ef}{$E_{\rm F}$}
\newcommand{\alalo}{$a_{\rm Al_2O_3}$}
\newcommand{\asto}{$a_{\rm STO}$}
\newcommand{\nst}{$N_{\rm STO}$}
\newcommand{\lnnlam}{(LNO)$_N$/(LAO)$_M$}
\newcommand{\lxolao}{(La$X$O$_3$)$_2$/(LaAlO$_3$)$_4$}
\newcommand{\xoalo}{($X_2$O$_3$)$_1$/(Al$_2$O$_3$)$_5$}

\title{Chern insulating phases and thermoelectric properties of EuO/MgO(001) superlattices}

\author{Okan K\"oksal}
\affiliation{Department of Physics and Center for Nanointegration Duisburg-Essen (CENIDE), University of Duisburg-Essen, Lotharstr. 1, 47057 Duisburg, Germany}
\author{Rossitza Pentcheva}
\email{Rossitza.Pentcheva@uni-due.de}
\affiliation{Department of Physics and Center for Nanointegration Duisburg-Essen (CENIDE), University of Duisburg-Essen, Lotharstr. 1, 47057 Duisburg, Germany}
\date{\today}

\begin{abstract}
The topological and thermoelectric properties of (EuO)$_{n}$/(MgO)$_{m}$(001) superlattices (SLs) are explored using density functional theory calculations including a Hubbard $U$ term together with Boltzmann transport theory. In (EuO)$_{1}$/(MgO)$_{3}$(001) SL at the lattice constant of MgO a sizable band gap of 0.51 eV is opened by spin-orbit coupling (SOC) due to a band inversion between occupied localized Eu $4f$ and $5d$ conduction electrons. This inversion between bands of opposite parity is accompanied by a spin reorientation in the spin-texture along the contour of band crossing surrounding the $\Gamma$ point and leads to a Chern insulator with $C$\,=\,--1, also confirmed by the single edge state. Moreover, this Chern insulating phase shows promising thermoelectric properties, e.g. a Seebeck coefficient between 400 and 800~$\mu$VK$^{-1}$. A similar SOC-induced band inversion takes place also in the ferromagnetic semimetallic (EuO)$_{2}$/(MgO)$_{2}$(001) SL. Despite the vanishing band gap, it leads to a substantial anomalous Hall conductivity with values up to --1.04 $e^{2}/h$ and somewhat lower thermoelectric properties. Both systems emphasize the relation between non-trivial topological bands and thermoelectricity also in systems with broken inversion symmetry. 
\end{abstract}

\maketitle

\section{Introduction}
\label{S:1}

Recently topological insulators (TIs), that are insulating in the bulk, but have topologically protected conducting edge states with dissipationless charge current at the surface, have  attracted a lot of attention in the field of condensed matter physics \cite{Zhang2010,Kane2010,Moore2010,Qi2011}. Several TIs from the V-VI group, e.g. Bi$_{2}$Se$_{3}$, Bi$_{2}$Te$_{3}$, Sb$_{2}$Te$_{3}$ are at the same time promising thermoelectric (TE) materials \cite{Zhang2009,Hsieh2009,Chen2009,Zhang2011} that can convert heat into electricity.  The connection and common characteristics between TI, in particular $Z_2$ TIs which preserve time-reversal symmetry (TRS), and TE such as heavy elements, narrow band gaps have been recently pointed out \cite{Muechler2013,Xu2017}. In contrast to our knowledge Chern insulators -- the TRS broken analogue of TIs -- have received little attention concerning TE applications. Chern insulators are characterized by a non-zero Chern number in the absence of an external magnetic field \cite{Weng2015,Ren2016}.  It is a significant challenge to realise robust Chern insulators. One strategy to achieve  breaking of TRS is via doping of magnetic impurities into known topological insulators, such e.g. Mn-doped HgTe-, Cr-, Fe-doped Bi$_{2}$Te$_{3}$, Bi$_{2}$Se$_{3}$, Sb$_{2}$Te$_{3}$ \cite{Liu_Zhang,Yu_Zhang,Fang_Bernevig}. Further realization possibilities are $5d$ transition metal atoms on graphene \cite{Zhang2012,Zhou_Liu} as well as OsCl$_3$ \cite{Sheng2017}.

\begin{figure}[htbp!]
\centering
\includegraphics[width=8.8cm,keepaspectratio]{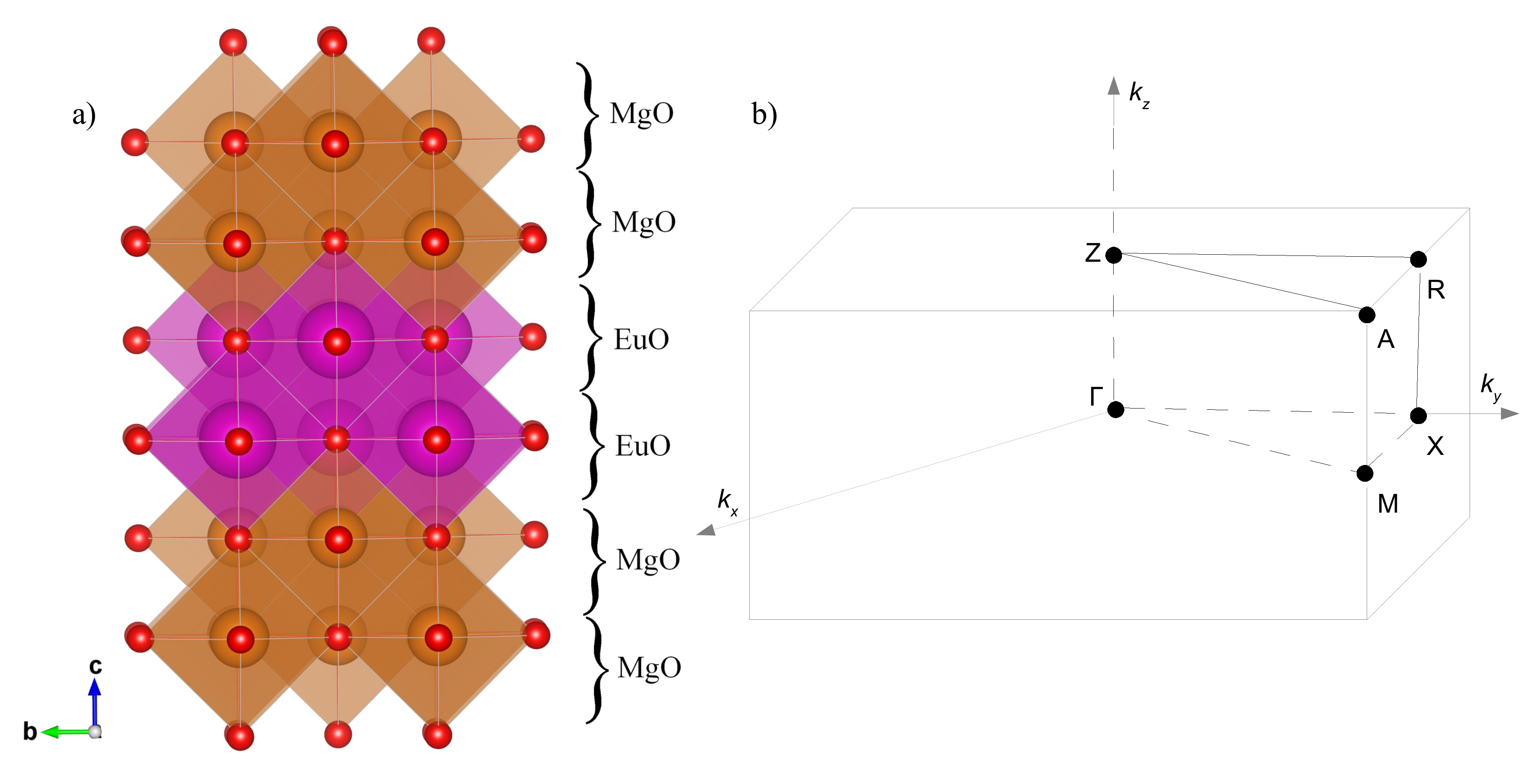}
\caption{a) Schematic view of the (EuO)$_{n}$/(MgO)$_{m}$(001) superlattice with $n=2$ and b) the corresponding Brillouin zone.} 
\label{fig:schematic_view}
\end{figure}

Transition metal oxides (TMO) with their rich functionality, resulting from the intricate interplay of spin, orbital and lattice degrees of freedom, have a greater tendency towards TRS breaking and larger band gaps compared to conventional $sp$ bonded systems and are thus an attractive class of materials to search for topologically non-trivial states. Intensive efforts have been directed at finding Chern insulators in TMO that host a honeycomb lattice, for which  Haldane predicted a quantized anomalous Hall effect in his seminal work for spinless fermions  \cite{Haldane}.  A buckled honeycomb lattice can be formed in (111)-oriented A$X$O$_3$ perovskite superlattices (SL) by two consecutive triangular $X$-layers as proposed by Xiao et al. \cite{Xiao2011}. Several realizations have been proposed, e.g. SrIrO$_3$ and LaAuO$_3$ bilayers, however, considering correlation effects results in an AFM ground state for SrIrO$_3$ \cite{Lado2013,Okamoto2014}. In the  $3d$ series, e.g. in (La$X$O$_3$)$_2$/(LaAlO$_3$)$_4$(111) SLs \cite{Doennig2016}, LaMnO$_3$ was identified as a Chern insulator with a band gap of ~150 meV when P321 symmetry is preserved, but the ground state is a trivial Mott insulator with Jahn-Teller (JT) distortion. Further candidates for quantum anomalous Hall insulators (QAHI) are LaRuO$_3$ and LaOsO$_3$ \cite{HongliNQM}, 
as well as the metastable symmetric ferromagnetic cases of LaPdO$_3$, LaPtO$_3$ and LaTcO$_3$ \cite{Guo2019,OKRP2019} honeycomb bilayers encased in LaAlO$_3$(111). Corundum-derived SL provide another realization of the honeycomb lattice. A systematic study of the $3d$ \cite{OKRP2018}, $4d$ and $5d$ \cite{JPCS2019} series of corundum-based honeycomb layers \xoalo(0001) identified the metastable cases of $X$\,=\,Tc,\,Pt as  Chern insulators with $C$\,=\,--2 and --1 and band gaps of 54 and 59 meV, respectively. 

Other lattice types proposed as candidate Chern insulators are e.g. rutile-derived heterostructures \cite{Huang_Vanderbilt,Cai_Gong,Lado2016}, pyrochlore \cite{Fiete2015}, as well as rocksalt-derived superlattices as EuO/CdO \cite{Zhang2014} and EuO/GdN SLs\cite{Garrity2014}. The aim in the latter is to combine heavy elements with large SOC in the two initially topologically trivial components in a quantum well (QW) structure and induce a SOC-dirven band inversion, analogous to the HgTe/CdTe QW, where band inversion was originally predicted and observed\cite{Bernevig2006,Zhang2006,Konig2007}. EuO is one of the few ferromagnetically (FM)  ordered semiconductors \cite{Mauger1986} with a Curie temperature ($T_c$) of 69 K. Similarly, in both EuO/CdO and EuO/GdN the CI state is achieved under considerable strain\cite{Zhang2014,Garrity2014}. In EuO/GdN the band inversion involves Eu $4f$  and the Gd $5d$\cite{Garrity2014}, whereas in EuO/CdO QW it takes place between the occupied Eu $4f$  and the Cd $5s$ states. The resulting band gap is in the meV range owing to the large $\Delta l\,=\,3$\cite{Zhang2014}. 

Here we follow a different strategy: by combining EuO with the large band gap insulator MgO in a (EuO)$_{1}$/(MgO)$_{3}$(001)  superlattice at the lattice constant of MgO (4.21 \AA) we achieve a band inversion in SL  between Eu $4f$ and $5d$ states within the same component, while MgO merely plays the role of a spacer.  We note that EuO/MgO(001) QW heterostructures have  already been realized experimentally\cite{Tjeng2009,Tjeng2011,Mueller2013,Mueller2016} using lattice-matched yttria-stabilized zirconia (YSZ).  

Already Hicks and Dresselhaus \cite{Hicks1993} proposed that the TE properties of materials can be improved in reduced dimensions as e.g. in QW. Experimentally, a giant Seebeck coefficient was reported in $\delta$-doped SrTiO$_3$ SLs\cite{Ohta2007}. The confinement- and strain-induced enhancement of TE properties was recently addressed based on first principles calculations  in \lno/\lao(001) SLs \cite{Benjamin2018,Benjamin2019,Viewpoint} and Sr$X$O$_3$/SrTiO$_3$(001) quantum wells \cite{Verma2019}, as well as other \sto-based SLs \cite{Pallecchi2015,Filippetti2012,Delugas2013,Bilc2016}. For example in (LaNiO$_3$)$_1$/(LaAlO$_3$)$_1$(001) \cite{Benjamin2018} the confinement- and strain-induced metal-to-insulator transition (MIT) at $a_{\rm STO}$ leads to enhanced in-plane power factor and high Seebeck coefficient. In the following we discuss the electronic properties of (EuO)$_{n}$/(MgO)$_{m}$(001) SL and provide topological analysis for the non-trivial cases. Moreover, we address in particular the implications of the topological Chern state in (EuO)$_{1}$/(MgO)$_{3}$(001) QWs on the thermoelectric properties using Boltzmann transport theory and compare to the (EuO)$_{2}$/(MgO)$_{2}$(001) case.

\section{Theoretical methods}
\begin{figure} [htp!]
\includegraphics[width=8.6cm,keepaspectratio]{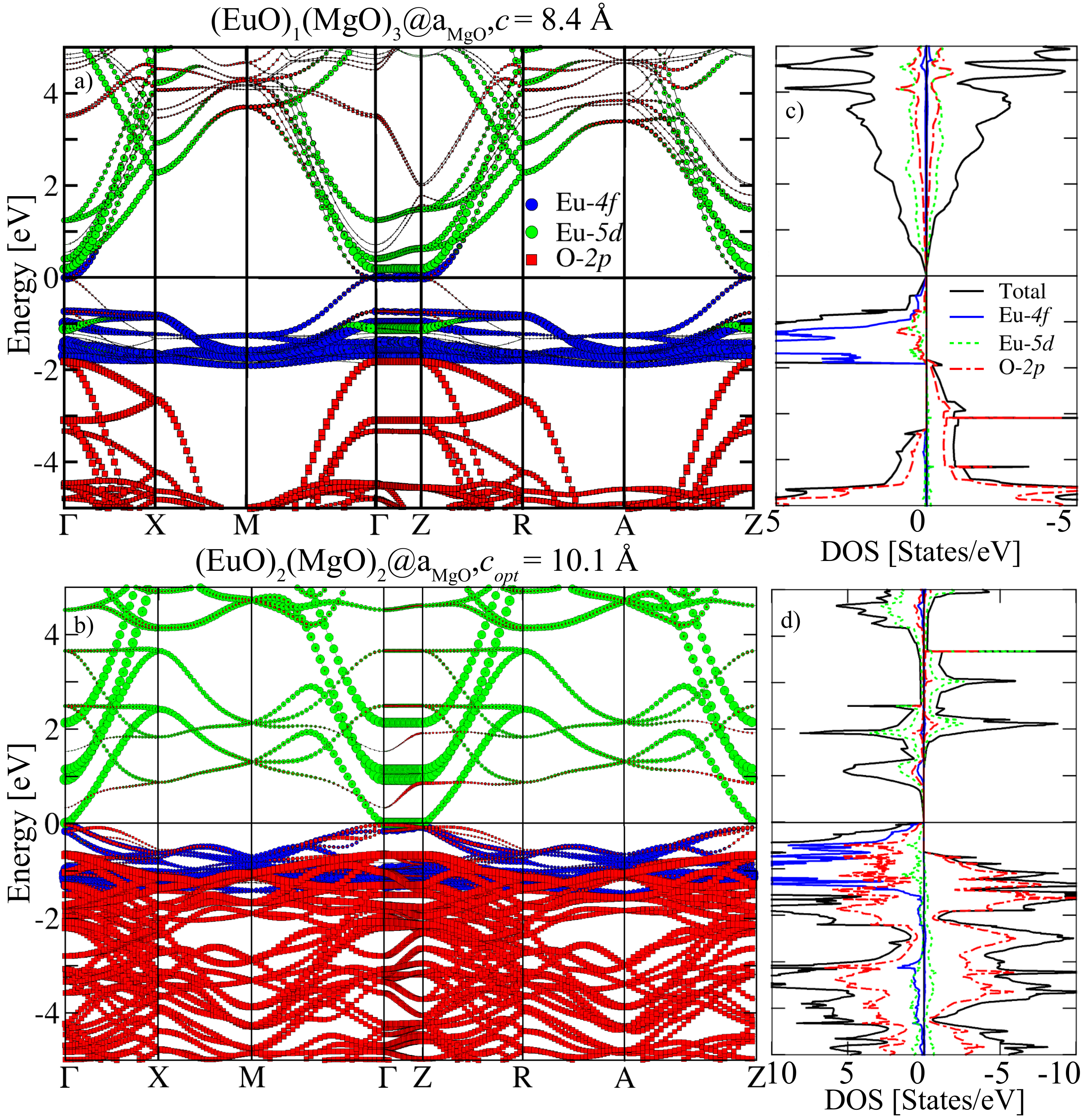}
\caption{a-b) Element- and orbital-projected band structure of ferromagnetic (EuO)$_{n}$/(MgO)$_{m}$(001) with an out-of-plane lattice parameter $c=2a_{\rm MgO}$ and optimized $c$ for $n=2$, $m=2$. The corresponding projected density of states (DOS) are displayed in c-d). The orbital character is color coded, highlighting  that the states below/above the Fermi level have predominantly Eu-$4f$ (blue)/Eu-$5d$ (green) as well as O-$2p$ (red) character, respectively.}
\label{fig:EuO1_MgO3}
\end{figure}

\begin{figure*} [htbp!]
\includegraphics[width=18cm,keepaspectratio]{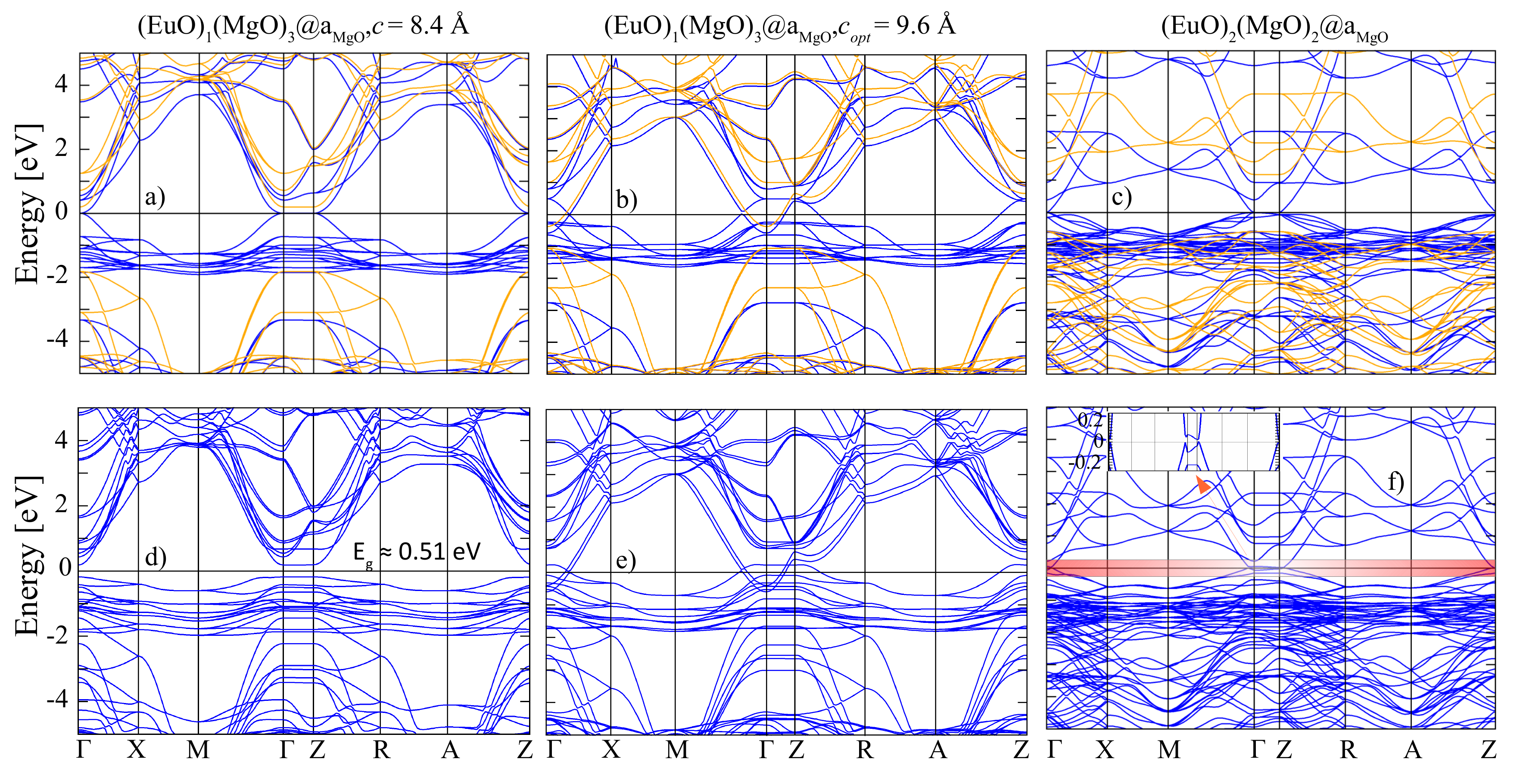}
\caption{GGA+$U$ band structures of ferromagnetic (EuO)$_{n}$/(MgO)$_{m}$(001) at fixed lateral lattice constant of MgO and a) constrained or b) optimized out-of-plane lattice constant $c$ for $n=1$, $m=3$, as well as c) optimized  $c$ for $n=2$, $m=2$. Majority and minority channels are shown in dark blue/light orange. The corresponding GGA+$U$+SOC band structures are displayed in the bottom panels d-f).}
\label{fig:EuO_MgO_BS}
\end{figure*}

\begin{figure*} [htp!]
\centering
\includegraphics[width=12cm,keepaspectratio]{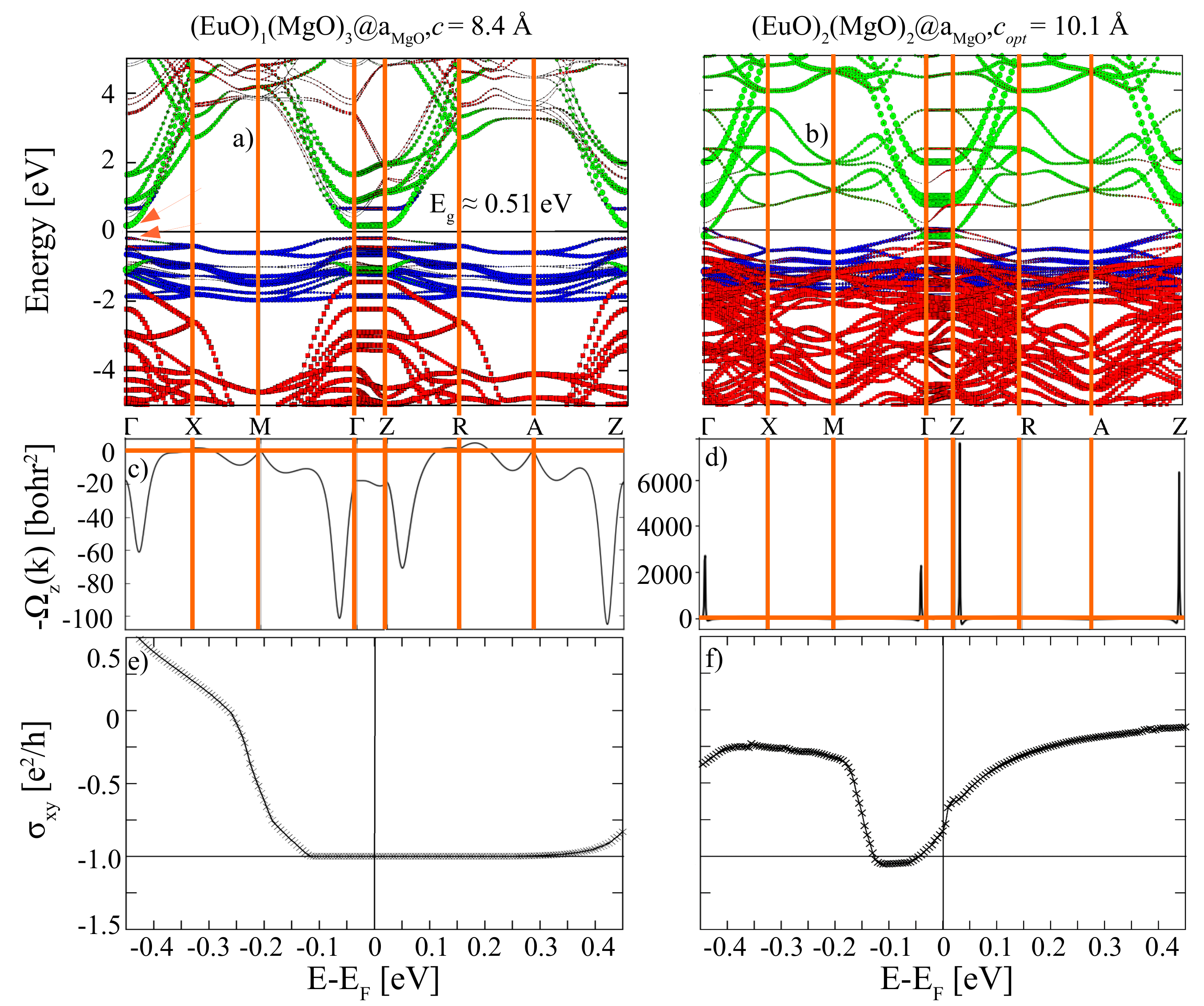}
\caption{In a-b) GGA+$U$+SOC band structures for FM (EuO)$_{1}$/(MgO)$_{3}$(001) and (EuO)$_{2}$/(MgO)$_{2}$(001) with magnetization along the [001]-direction as well as c-d) Berry curvatures along the same $k$-path. The corresponding anomalous Hall conductivities (AHC) $\sigma_{xy}$ in units of $e^{2}/h$ as a function of the chemical potential are displayed in e-f).}
\label{fig:EuO1_MgO3_SOC}
\end{figure*}

\begin{figure*} [htp!]
\includegraphics[width=18cm,keepaspectratio]{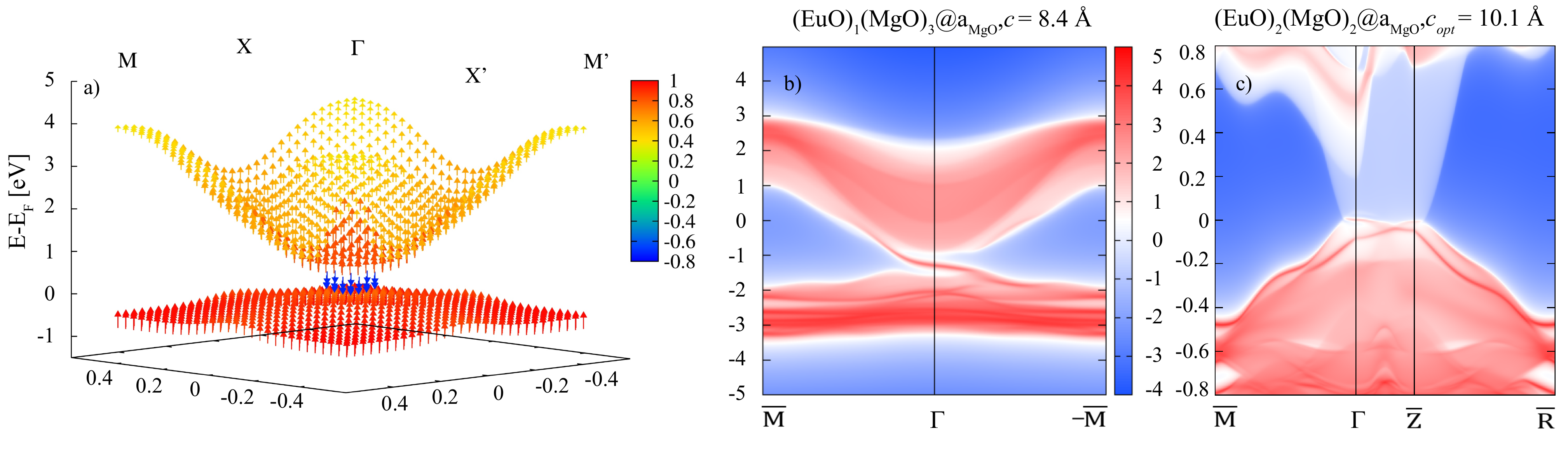}
\caption{a) The band-decomposed spin-texture in $k$-space from the GGA+$U$+SOC calculation with magnetization along [001] for the topmost occupied and lowest unoccupied bands (cf. Fig. \ref{fig:EuO1_MgO3_SOC}a) of the FM (EuO)$_{1}$/(MgO)$_{3}$(001) superlattice. The color scale denotes the projection on the $\hat{z}$-axis with red (blue) indicating parallel (antiparallel) orientation. The calculated edge states for (EuO)$_{1}$/(MgO)$_{3}$(001) and (EuO)$_{2}$/(MgO)$_{2}$(001) are shown in b-c) . Warmer colors (red/white) represent higher local DOS, blue regions denote the bulk energy gap and solid red lines are the edge states connecting the valence and conduction bands.}
\label{fig:EuO1_MgO3_ST_and_SS}
\end{figure*}

Density functional calculations were performed for (EuO)$_{n}$/(MgO)$_{m}$(001) SLs with the projector augmented wave (PAW) method\cite{PAW} as implemented in the VASP \cite{VASP} code. The cutoff energy of the plane-waves was set to 500 eV. For the exchange-correlation functional we used the generalized gradient approximation (GGA) by Perdew, Burke and Enzerhof \cite{GGA_PBE}. A $\Gamma$-centered $k$-point grid of 16$\times$16$\times$8 were adopted in the self-consistent calculations employing the tetrahedron method \cite{Bloechl1994}. Static electronic correlation effects were taken into account within the GGA\,+$U$ approach in the formulation of Liechtenstein et al. \cite{Liechtenstein}. Consistent with previous studies \cite{Larson2006,Larson2006PRB,Schlipf2013,Tong2014}, an on-site Coulomb repulsion parameter of $U$\,=\,7.4 eV and an exchange interaction parameter $J$\,=\,1.1 eV were considered for the Eu $4f$ states. With these values we obtain a band gap of 1.13 eV for antiferromagnetic coupling which is in very good agreement with the experimentally reported band gap of 1.12 eV \cite{Guentherodt1971} of room-temperature bulk EuO, whereas for the ferromagnetic ground state the band gap amounts to 0.65 eV. The optimized bulk lattice constant of EuO with ferro- and antiferromagnetic arrangement within GGA+$U$ is  $a$\,=\,5.184\,\AA\, and $a$\,=\,5.193\,\AA, respectively, slightly higher than the experimental value of $a$\,=\,5.141\,\AA\ \cite{Vleck1961}. Similarly, for bulk MgO GGA yields a bulk lattice parameter of $a$\,=\,4.24\,\AA, somewhat larger than the experimental lattice constant $a$\,=\,4.21\,\AA \cite{Roessler1967,Boer1998,Fei1999}. We note that within GGA the band gap of MgO is significantly underestimated (4.28 eV), compared to the experimental value of 7.83 eV \cite{Whited1973} and can be improved only by considering many body effects \cite{Fuchs2008}. Still the GGA MgO band gap is much larger than the one of the active material EuO, thus the phenomena in the heterostructure are determined by the confined EuO and not affected by the size of the band gap of MgO. The heterostructures were modeled at the experimental lateral lattice constant of MgO and internal parameters were relaxed until the Hellman-Feynman forces are less than 1 meV/\AA, while the $c$ lattice constant was either fixed at the value of bulk MgO or relaxed.  In the case of (EuO)$_{1}$/(MgO)$_{3}$(001) the topologically non-trivial case was also explored using the all-electron full-potential linearized augmented plane wave (LAPW) method as implemented in the Wien2k code \cite{wien2k}. In particular, the anomalous Hall conductivity (AHC) calculations were performed on a dense $k$-point mesh of 144$\times$144$\times$12 using the wannier90 code \cite{wannier90}. The transport coefficients based on input from the DFT calculations are obtained within the constant relaxation time approximation using the BoltzTraP code \cite{Boltztrap}.

\section{Results and discussion:}
\subsection{GGA+$U$ results for ({\rm EuO})$_{n}$/({\rm MgO})$_{m}$(001) quantum wells}  
\label{sec:EuO_MgO_electronic_properties}

In this Section we discuss the electronic properties of the (EuO)$_{n}$/(MgO)$_{m}$(001) superlattices, referred to as ($n$,$m$) in the following. According to Hund's rule, Eu$^{2+}$ exhibits a formal $4f^7$ high-spin configuration with a closed shell and  a large magnetic moment of $\sim 7.0$ \mub. The ferromagnetic state is the ground state. The GGA+$U$ element- and orbitally resolved band structure and the spin-dependent projected density of states (DOS) of (EuO)$_{1}$/(MgO)$_{3}$(001) with $c_{\rm MgO}$ are shown in Fig. \ref{fig:EuO1_MgO3}a and c. Just below the Fermi level the band structure is dominated by the narrow (bandwidth $\sim$ 2 eV) half-filled Eu $4f$ bands, whereas the conduction bands are strongly dispersive, e.g. along M-$\Gamma$-X and of prevailing Eu $5d$ character (see Fig. \ref{fig:EuO1_MgO3}a), respectively.  Moreover, the  top of the valence and bottom of the conduction band of this quantum well touch along $\Gamma$-Z, rendering the system semimetallic with predominant contribution of majority spin bands. The main effect of the $c$ relaxation is the enhanced dispersion and overlap of conduction and valence bands along $\Gamma$-Z (cf. Fig. \ref{fig:EuO_MgO_BS}b). On the other hand, the band structure of (EuO)$_{2}$/(MgO)$_{2}$(001) with relaxed $c$ (cf. Figs. \ref{fig:EuO1_MgO3}b and \ref{fig:EuO_MgO_BS}c) bears some similarities to $n=1$, $m=3$ at the MgO $c$ lattice constant, in particular, the touching flat conduction and valence bands along $\Gamma$-Z, however exhibits a much stronger overlap and hybridization between the Eu $4f$ and O $2p$ bands and a pronounced O $2p$ contribution along $\Gamma$-Z just below the Fermi level, visible also in the orbitally projected DOS in Figs. \ref{fig:EuO1_MgO3}c and d. 

\begin{figure*} [htp!]
\includegraphics[width=16.5cm,keepaspectratio]{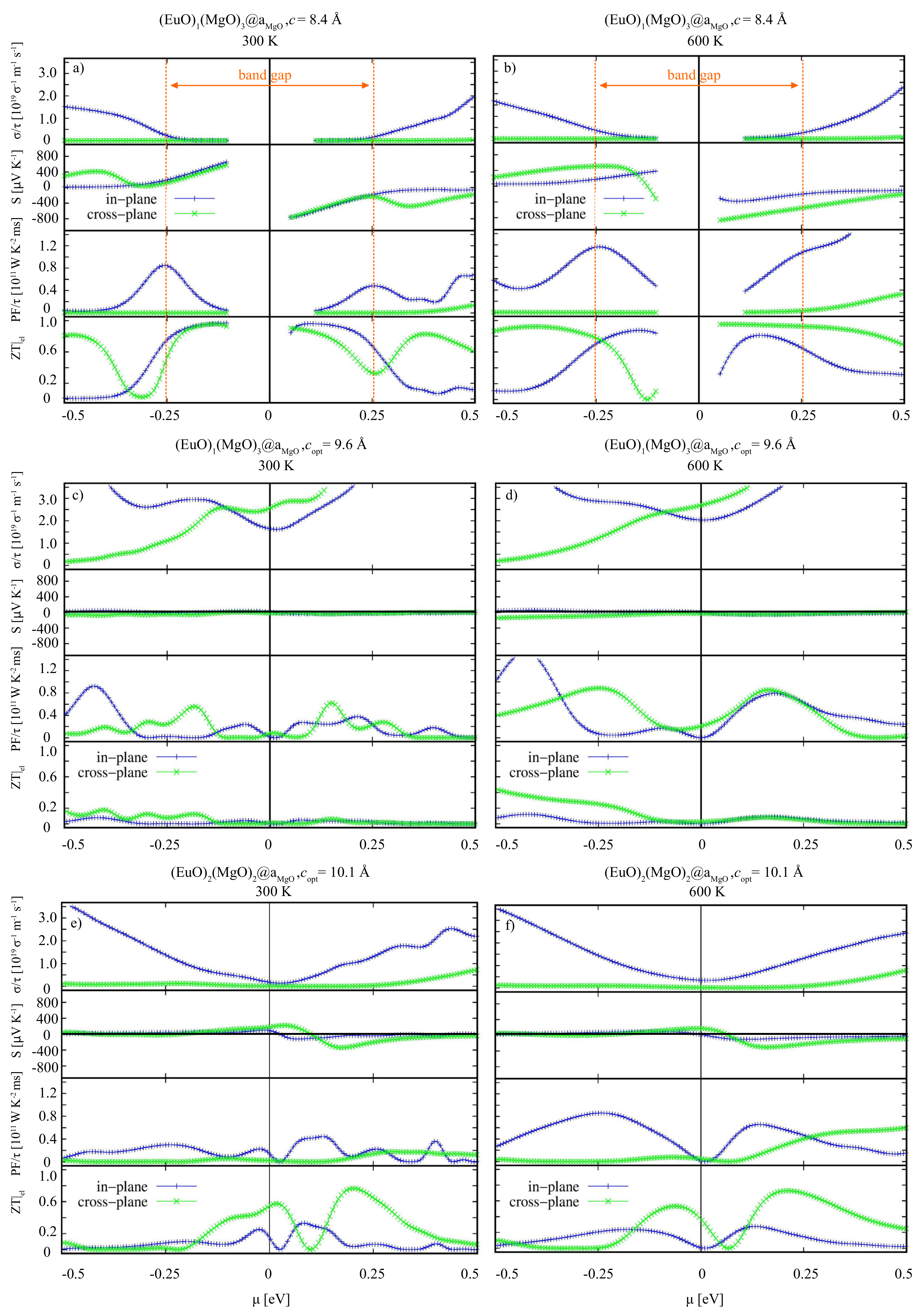}
\caption{The in- and cross-plane components of the electrical conductivity tensor divided by $\tau$, the Seebeck coefficient, $PF/\tau$ and the electronic contribution to the 
figure of merit ZT$|_{el}$ are shown as a function of the chemical potential in a-b) at 300 K and 600 K for FM (EuO)$_{1}$/(MgO)$_{3}$(001) superlattice at $c$\,=\,8.4~\AA, whereas c-d) show the  thermoelectric properties with the optimized out-of-plane parameter of $c_{opt}$\,=\,9.6~\AA, e-f) thermoelectric performance of (EuO)$_{2}$/(MgO)$_{2}$(001) SL.}
\label{fig:EuO1_MgO3_TE}
\end{figure*}

\subsection{Effect of spin-orbit coupling and topological analysis}
\label{sec:CI_phase}

Despite the similar features in the bandstructure, the effect of spin-orbit coupling is very distinct for the three systems. The corresponding band structures are displayed in Fig. \ref{fig:EuO_MgO_BS}d-f. While (EuO)$_{1}$/(MgO)$_{3}$(001) with relaxed $c$  remains metallic with no pronounced rearrangement of bands, for (EuO)$_{1}$/(MgO)$_{3}$(001) at $c=2a_{\rm MgO}$ a significant band gap of 0.51 eV is opened for SOC with out-of-plane magnetization direction. 
Apparently, the degeneracy of the touching bands at the Fermi level is lifted giving rise to a band inversion along $\Gamma$-Z. The band inversion is present but the band gap is nearly vanishing for (EuO)$_{2}$/(MgO)$_{2}$(001) with relaxed $c$ (see inset in Fig. \ref{fig:EuO_MgO_BS}f). In order to analyze the origin of the band rearrangement and inversion we plot in Fig. \ref{fig:EuO1_MgO3_SOC}a and b the element and orbital projections on the band structure with SOC for (1,3) and (2,2).  
In contrast to the previously reported band inversion between Eu $4f$ and Cd $5s$ bands in EuO/CdO(001) \cite{Zhang2014} or Eu $4f$ and Gd $5d$ states in EuO/GdN SL \cite{Garrity2014}, for(EuO)$_{1}$/(MgO)$_{3}$(001) at $c=2a_{\rm MgO}$ the band inversion takes place between the $4f$ and $5d$ states of Eu itself. The strong interaction of these bands of opposite parity and $\Delta l\,=\,1$ lead to a substantial band gap opening. Interestingly this bears analogies with previous reports of bulk EuO under pressure, where fluctuations between $(4f)^{7}(5d)^{0}$ and $(4f)^{6}(5d)^{1}$ configurations were suggested in experimental \cite{Zimmer1984} and theoretical studies \cite{Eyert1986PB,Eyert1986SS}. Upon inclusion of SOC the band structure of (EuO)$_{2}$/(MgO)$_{2}$(001) with relaxed $c$ shows a reduced contribution of O $2p$ along $\Gamma$-Z and a similar inversion of the topmost Eu $4f$ and lowest $5d$ band around \ef, though with a vanishing band gap.

Having identified the origin of the band rearrangement and inversion for the two systems, we proceed to analyze the topological properties. The non-trivial nature of the (1,3) system is underpinned by the Berry curvature (cf. Fig. \ref{fig:EuO1_MgO3_SOC}c) which exhibits pronounced negative peaks along $\Gamma$-Z as well as M-$\Gamma$ paths, and a flat region along $\Gamma$-Z. This leads to the emergence of a broad plateau in the anomalous Hall conductivity $\sigma_{xy}$ at \ef\ in Fig. \ref{fig:EuO1_MgO3_SOC}e, rendering (EuO)$_{1}$/(MgO)$_{3}$(001) a Chern insulator with a quantized $C$\,=\,--1. As shown in Fig. \ref{fig:EuO1_MgO3_SOC}d, sharp peaks arise in the Berry curvature $\Omega(k)$ of (EuO)$_{2}$/(MgO)$_{2}$(001) at the avoided crossing of Eu $4f$ and $5d$ bands along the M-$\Gamma$ and Z-R paths with values of 3000 and 8000 bohr$^{2}$. 
For (EuO)$_{2}$/(MgO)$_{2}$(001) the Hall conductivity in Fig. \ref{fig:EuO1_MgO3_SOC}f shows substantial, nearly quantized values (--1.04 $e^{2}/h$) caused by the non-trivial bands with the plateau being just below \ef\ and a finite value of 0.8\,$e^{2}/h$ at \ef.

In Fig. \ref{fig:EuO1_MgO3_ST_and_SS}a we plot the spin-texture of the relevant bands of (1,3)  (see Fig. \ref{fig:EuO1_MgO3_SOC}a). The occupied band exhibits only positive $s_z$ values throughout the whole BZ. In contrast, the out-of-plane spin component $s_z$ of the lower part of the unoccupied parabolic band is negative  around $\Gamma$ but reverses sign further away from the BZ center. This switching of  spin orientation can be ascribed to the SOC-induced band inversion between the occupied majority $4f$ states and unoccupied minority $5d$ states band along $\Gamma$-Z, occurring just above \ef\ in Fig.~\ref{fig:EuO_MgO_BS}a. The surface states shown in Fig.~\ref{fig:EuO1_MgO3_ST_and_SS}b using Wanniertools \cite{wanniertools} based on the Maximally Localized Wannier functions (MLWF) method presents one topologically protected chiral edge state, connecting the valence and conduction band. In contrast, the edge state for (EuO)$_{2}$/(MgO)$_{2}$(001) in Fig.~\ref{fig:EuO1_MgO3_ST_and_SS}c is obscured due to the overlap of valence and conduction bands along $\Gamma$-Z.

\subsection{Thermoelectric properties}
\label{sec:thermoelectric}

In the following, we investigate the thermoelectric properties of the (EuO)$_{n}$/(MgO)$_{m}$(001) SLs with and without optimized out-of-plane lattice constants $c$. A central quantity related to the TE efficiency is the figure of merit:

\begin{equation} 
\label{eq:1}
ZT = \frac{S^{2} \sigma}{\kappa} T				
\end{equation}

where $S$ is the Seebeck coefficient, $\sigma$ the conductivity and $\kappa$ the thermal conductivity. Another related quantity is the power factor $PF=S^2\sigma$. In Fig. \ref{fig:EuO1_MgO3_TE} we plot $\sigma/\tau$, the Seebeck coefficient and $PF/\tau$ for the systems studied in Fig. \ref{fig:EuO_MgO_BS} at two different temperatures, 300 and 600 K. While (EuO)$_{1}$/(MgO)$_{3}$(001) with $c$\,=\,8.4~\AA\ is a Chern insulator upon inclusion of SOC (cf. Fig. \ref{fig:EuO1_MgO3_TE}), the other two systems remain metallic or semimetallic. This is reflected in the vanishing out-of-plane conductivity for the former, whereas the more dispersive bands of (EuO)$_{1}$/(MgO)$_{3}$(001) with relaxed $c$ leads to higher $\sigma/\tau$. On the other hand the flat touching bands along $\Gamma$-Z in (EuO)$_{2}$/(MgO)$_{2}$(001) result in a vanishing out-of-plane conductivity.  
The Chern insulating system (EuO)$_{1}$/(MgO)$_{3}$(001) exhibits a much higher Seebeck coefficient, reaching values between 400-800~$\mu$VK$^{-1}$ and $PF/\tau$ of 0.8$\cdot$10$^{11}$ (300 K) and 1.2$\cdot$10$^{11}$ W/K$^{2}$ms (600 K). The electronic figure of merit ZT$|_{el}$ attains in- and out-of-plane values of $\sim 0.8$  and $\sim 0.5$ at 300 K and $\sim 0.8$  and $\sim 0.7$ at 600 K at the valence band edge while values of $\sim 0.4$ and $\sim 0.6$ at 300 K and $\sim 0.9$ and $\sim 0.7$ at 600 K at the conduction band edge are obtained, respectively. In contrast $S$, $PF/\tau$ and electronic figure of merit ZT$|_{el}$ are much lower for the metallic case with optimized $c$. On the other hand, (EuO)$_{2}$/(MgO)$_{2}$(001) SL exhibits significant values for $S$ ($\sim 200$ $\mu$VK$^{-1}$), $PF/\tau$ (0.8$\cdot$10$^{11}$ W/K$^{2}$ms at 600 K) and ZT$|_{el}$ ($\sim 0.6$) may be reached upon doping.


Thus, both systems with topologically non-trivial bands and in particular the Chern insulating phase show promising TE properties. The improved performance is associated with the mixture of flat bands around $\Gamma-$Z due to the SOC-induced band inversion that leads to a steep increase of DOS at the band edges, thereby enhancing $S$ and dispersive bands that contribute to the electrical conductivity. Moreover, while we consider here only the electronic contributions to the thermal conductivity, we expect that the high atomic number of Eu (63) and the phonon scattering at interfaces in this layered structure will be beneficial to reduce the lattice contribution to $\kappa$.

\section{Summary}
\label{sec:fin}
In summary, the effect of confinement and strain on the topological and thermoelectric properties of (EuO)$_{n}$/(MgO)$_{m}$(001) superlattices has been studied by DFT\,+\,\textit{U}\,+\,SOC calculations in conjunction with the semi-classical Boltzmann transport theory. Combining two topologically trivial materials EuO and MgO in a QW structure results in a Chern insulating phase. Particularly, (EuO)$_{1}$/(MgO)$_{3}$(001) SL with lattice parameters constrained to the ones of MgO exhibits semimetallic behavior. The inclusion of SOC opens a large band gap of 0.51 eV due to a band inversion between Eu $5d$ and $4f$ bands. This mechanism is distinct to the one in EuO/CdO \cite{Zhang2014} and EuO/GdN SLs\cite{Garrity2014}, where the band inversion takes place between bands of different elements in the two constituents: Eu $4f$ and Cd $5s$ or Gd $5d$, respectively. The resulting Chern insulating phase with $C$\,=\,--1 shows a sign reversal of the out-of-plane spin components $s_z$ along the loop of band inversion around $\Gamma$ and a single chiral edge state. A similar band inversion occurs also in (EuO)$_{2}$/(MgO)$_{2}$(001) SL but with a vanishing band gap. The resulting band rearrangement close to \ef\ leads to sharp peaks of the Berry curvature at the avoided band-crossing along Z-R and a plateau in the anomalous Hall conductivity above \ef\ with a notable value of --1.04 $e^{2}/h$.  
Moreover, the (EuO)$_{1}$/(MgO)$_{3}$(001) SL exhibits enhanced thermoelectric performance in terms of Seebeck coefficient of 400-800~$\mu$VK$^{-1}$ and $PF/\tau$ of 0.8-1.2$\cdot$10$^{11}$W/K$^{2}$ms  and an electronic figure of merit of 0.4-0.9 at the band edges depending on temperature, driven by both the confinement and topological nature of the system. Similarly, an out-of-plane electronic $ZT$ of 0.6 is achievable in (EuO)$_{2}$/(MgO)$_{2}$(001) SL. 
The opening of a gap due to SOC-driven band inversion with steep increase of DOS at the band edges due to the flatness of bands along the $\Gamma$-Z direction promotes high values of the Seebeck coefficient and PF, whereas concomitant dispersive bands contribute to the electrical conductivity. Similar effect of systems at the verge of a metal-to insulator transition (though not topological) has been found in other oxide heterostructures \cite{Benjamin2018, Verma2019}. Furthermore, the results establish a link between topological and thermoelectric properties, in particular for systems with broken inversion symmetry.

\clearpage
\begin{acknowledgments}
We acknowledge useful discussions with M. M\"uller and A. Lorke on EuO/MgO quantum wells and funding by the German Science Foundation (Deutsche Forschungsgemeinschaft, DFG) within the CRC/TRR80, project G3 and computational time at the Leibniz Rechenzentrum Garching, project pr87ro. We also would like to thank G\'{e}rald K\"ammerer for performing initial calculations with Wien2k. 
\end{acknowledgments}

\end{document}